\documentstyle[preprint,epsf,aps]{revtex}
\tighten
\begin{document}
\title{\bf Classical Chaos and its Quantum Manifestations}
\author{Guest Editors \\ $\;$\\
J.~Bellissard$^{(a)}$, O.~Bohigas$^{(b)}$, G.~Casati$^{(c)} $
and D.~L.~Shepelyansky$^{(a)}$}

\address{(a)Laboratoire de Physique Quantique, UMR 5626 du CNRS, 
Universit\'e Paul Sabatier,\\ 31062 Toulouse Cedex, France \\
(b)Division de Physique Th\'eorique, Institut de Physique Nucl\'eaire,\\
91406, Orsay Cedex, France\\
(c)International Center for the Study of Dynamical Systems,\\
via Lucini, 3, I-22100 Como, Italy
}

\date{29 March, 1999}
\maketitle

\begin{abstract}
We present here the special issue of Physica D
in honor of Boris Chirikov. It is based on the proceedings
of the Conference {\it Classical Chaos and its Quantum Manifestations}
held in Toulouse in July 1998. This electronic version
contains the list of contributions, the introduction
and the unformal conclusion. The introduction represents
{\it X Chirikov Chaos Commandments} and reviews Chirikov's
pioneering results in the field of classical and quantum chaos.
The conclusion, written by P.M.Koch,
gives an outlook on the development of the field 
of chaos associated with Chirikov,
including the personal reminiscences of
Professor Andy Sessler. This electronic version 
of the special issue has certain differences and extentions
comparing to the journal.

\end{abstract}

\newpage
\begin{center}
{\huge\bf Classical Chaos 

and its Quantum Manifestations}

{\large\bf Sputnik Conference of STATPHYS~20}

\parskip 0.3cm

{\Large\bf In honor of Boris Chirikov}

{\large\bf Toulouse, France $\ \bullet\ $ July 16 -- 18, 1998}

\end{center}

\epsfxsize=2.8in
\epsfysize=2.8in
\vskip -0.8cm
\hskip -1cm
\epsffile{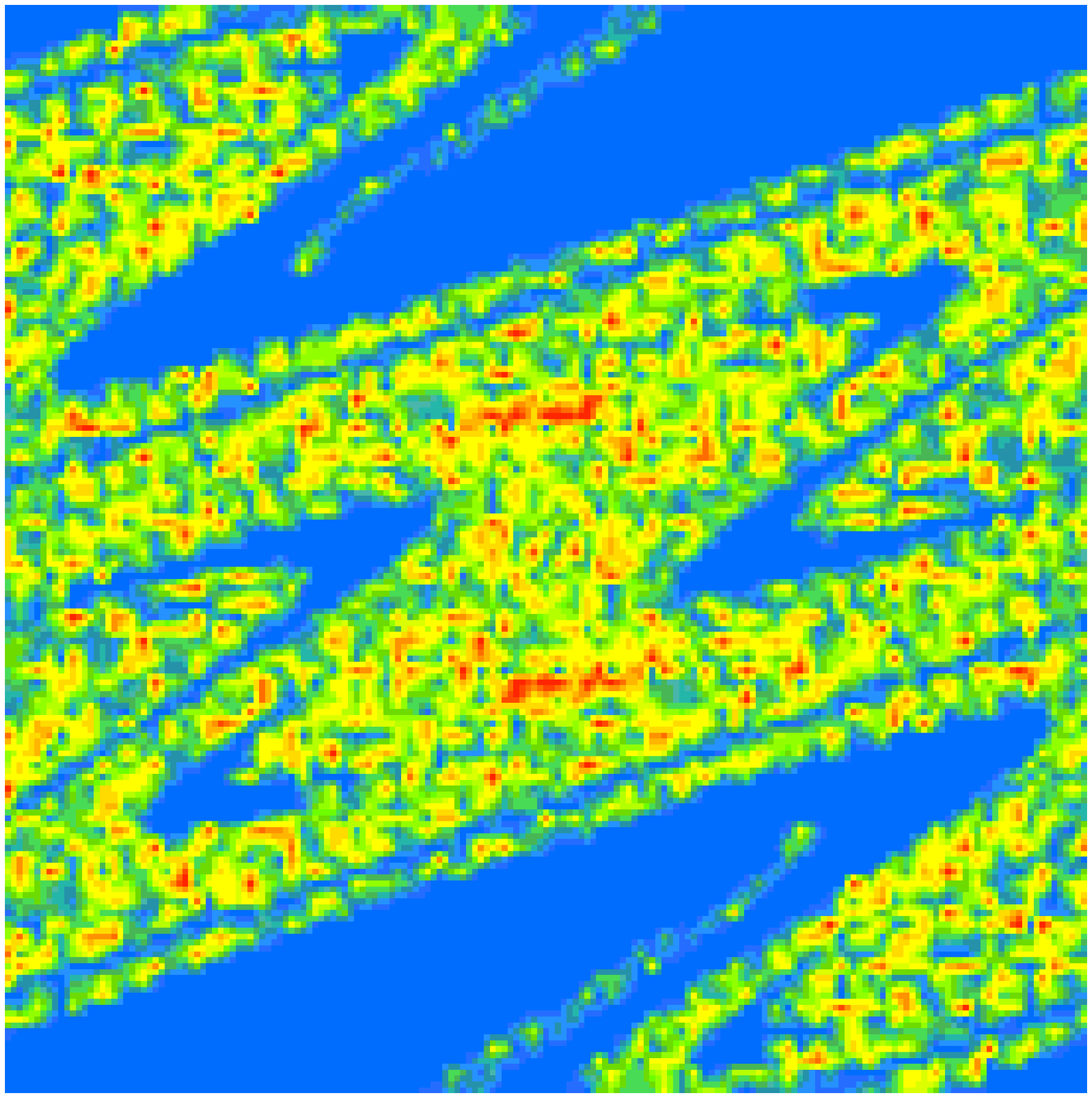}

\parskip 0.0 cm

\vskip -2.4in
\hskip +3.8in
{\large \underline{Organizers:} }
\medskip

\hskip +2.7in
{\large\bf Jean Bellissard,\quad Dima Shepelyansky}
\smallskip

\hskip +3in
(Universit\'e Paul Sabatier/CNRS, Toulouse)
\vspace{0.5cm}

\hskip +3.5in
{\large \underline{Scientific\phantom{g}committee:}}
\medskip

\hskip +2.7in
{\large {\bf J. Bellissard} (Toulouse)}
\smallskip

\hskip +2.7in
{\large {\bf O. Bohigas} (Orsay)}
\smallskip

\hskip +2.7in
{\large {\bf G. Casati} (Como)}
\smallskip

\hskip +2.7in
{\large {\bf D. Shepelyansky} (Toulouse)}
\vspace{0.7cm}

\begin{tabular}{lp{19cm}}
\hspace*{-1.5cm}  &
During the last decades, important mathematical progress in dynamical 
chaos has found applications to a broad range of physical systems. 
Recently, the manifestations of classical chaos have also become important 
in quantum systems, due to a rapid technological progress
in nanostructures and laser physics. 
The quantum and classical theory of chaos is developing
rapidly with applications to Rydberg atoms in external 
fields, cold atoms, mesoscopic systems, plasma physics, 
accelerators and planetary motion. 
The aim of the conference is to invite specialists from different 
fields to discuss the modern problems of classical and quantum chaos.
The conference is dedicated to the 70th birthday of Boris Chirikov,
who has contributed pioneering results in this area since 1959. 
\end{tabular}

\begin{center}
{\large\bf {\hskip -3cm} Invited speakers:}
\end{center}

{
\hskip -0.5cm
\begin{tabular}{lll}
V.Akulin (Paris)            & Y.Fyodorov (St. Petersburg)\ \ \ & M.Raizen (Austin) \\
S.Aubry (Saclay)            & T.Geisel (G\"ottingen)      & S.Ruffo (Florence)   \\
E.Bogomolny (Orsay)         & I.Guarneri (Como)           & T.Seligman (Cuernavaca) \\
O.Bohigas (Orsay)           & F.Haake (Essen)             & E.Shuryak (Stony Brook) \\
A.Buchleitner (Munich)      & F.Izrailev (Novosibirsk)    & Y.Sinai (Princeton)  \\
L.Bunimovich (Atlanta)      & P.Koch (Stony Brook)        & A.D.Stone (Yale)  \\
G.Casati (Como)             & J.Laskar (Paris)            & F.Vivaldi (London) \\
B.Chirikov (Novosibirsk)\ \ \ & A.Lichtenberg (Berkeley)  & G.Zaslavsky (New York)  \\
S.Fishman (Haifa)           & R.MacKay (Cambridge)        & \\
V.Flambaum (Sydney)         & A.Pikovsky (Potsdam)       & {\em (participation confirmed)} \\
\end{tabular}
}
\medskip

{ A poster session will be organized for other participants.}

\medskip

\hskip -1.5 cm
{\large {\bf Sponsors: } CNRS, Universit\'e Paul Sabatier, 
Institut Universitaire de France, 

\hskip 1.3cm IRSAMC,\ \ R\'egion Midi--Pyr\'en\'ees,\ \  Physica D}
\bigskip

\hskip -1.5 cm
{\large {\bf Contact address:} Sylvia Scaldaferro (Registration), 
Dima Shepelyansky, } 

\hskip -1.5 cm Laboratoire de Physique Quantique, 
UMR du CNRS 5626, IRSAMC, Universit\'e Paul Sabatier, 

\hskip -1.5 cm 118, route de Narbonne, F-31062 Toulouse Cedex 4, France 

\hskip -1.5 cm Tel. +33 5 6155 6834\ \ \ Fax : +33 5 6155 6065 \ \ \ 
WWW: http://w3-phystheo.ups-tlse.fr/conf98/

\hskip -1.5 cm Email: sylvia@irsamc2.ups-tlse.fr, dima@irsamc2.ups-tlse.fr


\newpage
\textheight 24truecm
\textwidth 17truecm
\oddsidemargin -0.6truecm
\evensidemargin 0truecm
\topmargin 0cm
\topskip 0cm
\voffset -1.5cm
\parskip 12pt

\newpage

\epsfxsize=3.0in
\epsfysize=4.0in
\vskip 4.0cm
\hskip 3.0cm
\epsffile{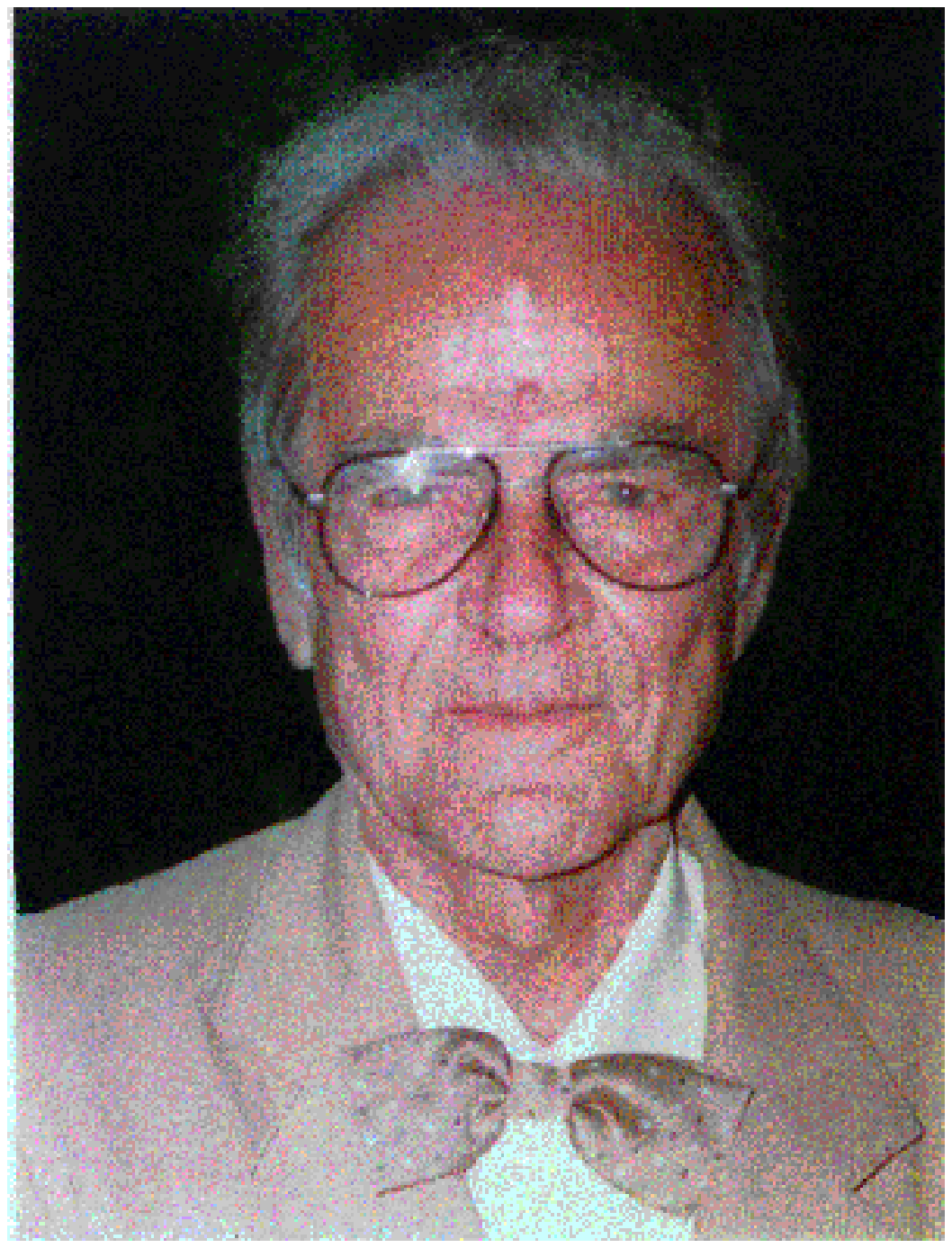}

\begin{center}
{\large\bf {\hskip -3cm} Boris Chirikov}\\
\end{center}
\begin{center}
{\bf {\hskip -3cm} 6 June, 1998}
\end{center}

\newpage

\bibliography{ref}
\vskip 1.0cm
\begin{center}
{\Large \bf Sputnik of Chaos}
\end{center}

\vspace{1cm}

\indent
This special issue of Physica D represents selected proceedings
of the Sputnik Conference of STATPHYS 20 ``Classical Chaos
and its Quantum Manifestations'' organized in honor of
Boris Chirikov, on the occasion of his 70th birthday on June 6th,
and held in Toulouse, France  on 16--18 July, 1998.

\indent
{\it Sputnik\/} is  the name of the very first artificial satellite, 
launched back in 1957. 
In russian, this word means companion, but it is also
a metaphor for a pioneering, outstanding achievement. 
These meanings seem appropriate for a portrait of Boris Chirikov. 
Back in 1959, he published a seminal article \cite{1959}, 
where he introduced the very first physical criterion for the onset 
of chaotic motion in deterministic Hamiltonian systems.
He then applied such criterion 
---now known as the {\em Chirikov resonance-overlap criterion}---
to explain puzzling experimental results on plasma confinement 
in magnetic bottles. As in an old oriental tale, Boris opened 
such a bottle, and freed the genie of Chaos, which spread the world over.

Boris Chirikov's research on chaos begun in a laboratory at the 
Kurchatov Institute for Atomic Energy (Moscow). 
In September 1959, he moved to Novosibirsk, at the Institute 
of Nuclear Physics founded by G.~I.~Budker, where he still continues to work.
He became a correspondent member of the Russian Academy of 
Sciences in 1983, and a full member in 1992.
(Further biographical details can be be found in \cite{aufn}.)

Giving a fair account of Chirikov's scientific output is an 
arduous task. Below we shall list ten prominent achievements 
of his, selected among those which are closer to the theme of this 
Conference, but we do so aware that much will be necessarily left out. 
What is more difficult to convey is a flavour of his personal
qualities, his warmth and kindness, his attitude on life and 
science, and the influence these had on our scientific community. 
An event in his life paints some traits of his character: 
as a young researcher, he left Russia's capital city, the 
hectic rush for career and influence, and chose a simpler 
way of life, in a remote Siberian forest.

\newpage
\epsfxsize=6.5in
\epsfysize=9.0in
\vskip -0.8cm
\hskip -1cm
\epsffile{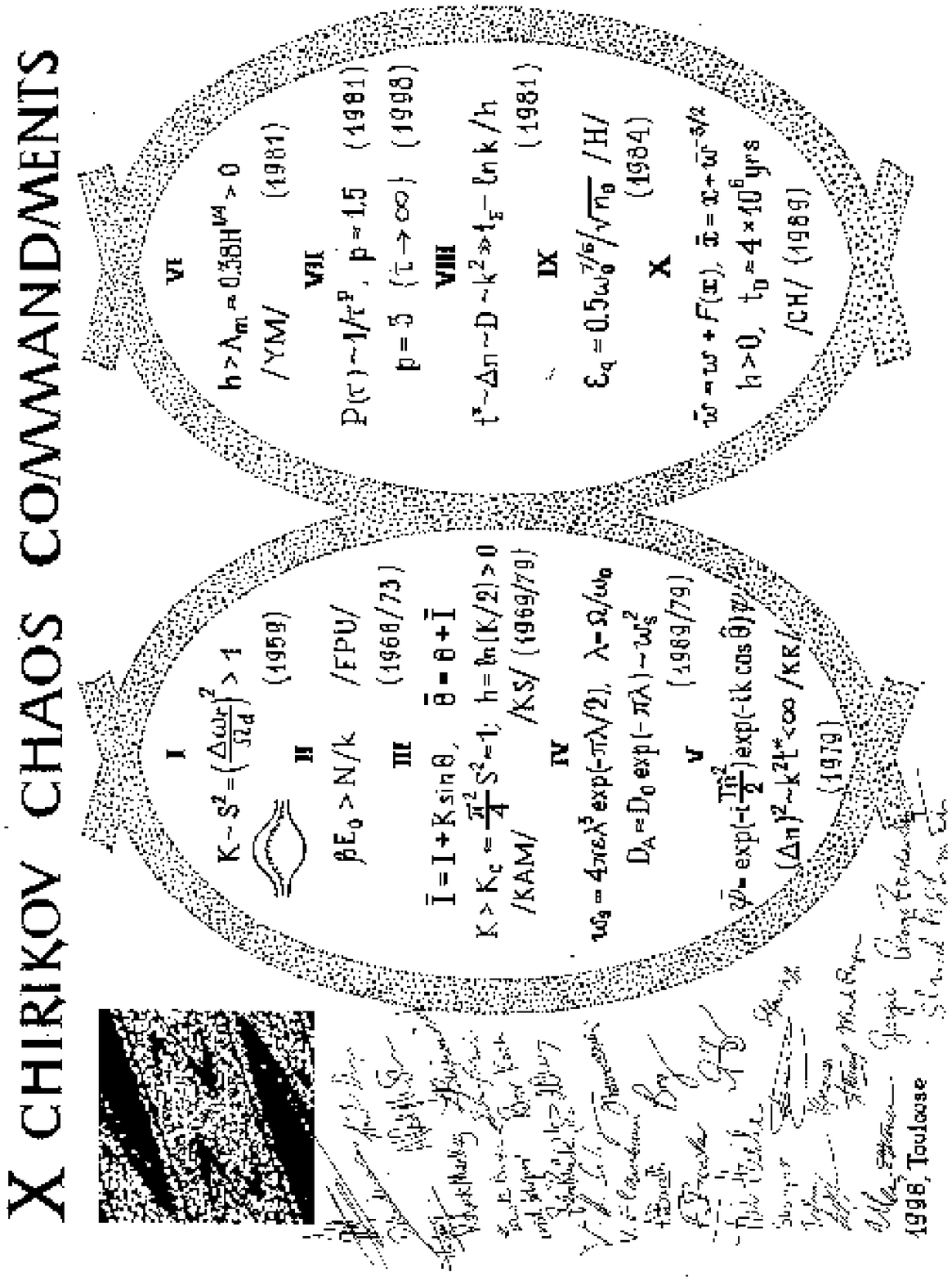}

\newpage
\hskip 5.0cm {\large \bf {\em   X  Chirikov Chaos "Commandments"}}
\footnote{ ) According to the Merriam-Webster dictionary
the word {\it "commandment"} means command, order. We use this word
as an equivalent  of order, law, rule, result, postulate, principle etc.
Nevertheless, a reader can note that even a strict rule (law)
can lead to chaotic unpredictable behavior in agreement with
the theory of deterministic chaos (see \cite{lich} for details).}

\indent
{\it I.} The first rule represents 
the Chirikov resonance-overlap criterion
introduced in \cite{1959} and then successfully applied 
to the determination of the confinement border for Rodionov
experiments \cite{rodion} with plasma in open mirror traps 
(the shape of magnetic lines is shown in the Figure). 
According to this criterion, a deterministic trajectory will begin to 
move between two nonlinear resonances in a chaotic and unpredictable manner,
in the parameter range $K \sim S^2 >1$. Here $K$ is the perturbation parameter,
while $S = \Delta\omega_r/\Omega_d$ is the resonance-overlap parameter,
given by the ratio of the unperturbed resonance width in frequency 
$\Delta \omega_r$ (often computed in the pendulum
approximation and proportional to the square-root of perturbation),
and the frequency difference $\Omega_d$ between two unperturbed resonances. 
Since its introduction, the Chirikov criterion has become 
an important analytical
tool for the determination of the chaos border \cite{cnote}.
The accuracy of the criterion can be improved on the
basis of a renormalization approach to resonances on smaller
and smaller scales \cite{escande}.
For an up-to-date account of the status of particle confinement in
magnetic traps, see \cite{plasma,london,open}.

\indent
{\it II.} This result \cite{chifpu} determines  the energy border
for strong chaos in the Fermi-Pasta-Ulam problem (FPU) \cite{fpu},
which became a cornerstone in modern statistical mechanics
(see the historical review in \cite{ford}).
The system represents a chain of $N$ weakly coupled nonlinear
oscillators with the
Hamiltonian  $H={\sum_n} [p_n^2/2+(x_{n+1} - x_n)^2/2 +
\beta (x_{n+1} - x_n)^4/4]$; initially only few long wave modes
with wave vector $k$ and energy $E_0$ are excited. The
chaos border is obtained from the Chirikov resonance-overlap criterion.
New insights on this problem can be found in
\cite{fpu1,fpu2}.

\indent
{\it III.} Here the first line represents the
Chirikov standard map \cite{1969,1979}. It is an area-preserving map
with action variable $I$ and phase $\theta$. The bars denote the new 
values of the variables, while $K$ is the perturbation parameter. 
The dynamics becomes unbounded and diffusive in $I$ for 
$K > K_c \approx 1$, when all Kolmogorov-Arnold-Moser (KAM) 
invariant curves are destroyed. This value is obtained
from the overlap criterion of first and higher order resonances 
\cite{1969,1979}. For large $K > 4$ the Kolmogorov-Sinai (KS)
entropy $h$, related to the exponential local instability of motion,
is well-described by the given analytical formula.
The Chirikov standard map provides a local description of the
interaction between resonances, which finds applications in 
such diverse physical systems as particles in magnetic traps 
\cite{1979}, accelerator physics \cite{felix}, highly excited 
hydrogen atoms in a microwave field \cite{dec2}, mesoscopic 
resonance tunnelling diode in a tilted magnetic field \cite{rtd}.
Later, a refined analysis \cite{greene,mackay} gave the
more precise value $K_c = 0.9716...$, related to the 
destruction of the KAM curve with the golden rotation number. 
For $K > K_c$ invariant curves are replaced by "cantori".
Rigorous results were obtained in \cite{aubry,mather,percival}
(see also \cite{mather2}).
However, in spite of fundamental advances in ergodic theory 
\cite{sinai}, a rigorous proof of the existence of a set of 
positive measure of orbits with positive entropy is still 
missing, even for specific values of $K$. 

\indent
{\it IV.} Here the first line represents the width $w_s$ of 
chaotic layer appearing around a separatrix of a nonlinear
resonance \cite{1969,1979}. This answers a question first
addressed by Poincar{\'e} \cite{poincare},
who estimated the angle of separatrix splitting, 
but not the width of the layer. The above equation is written
for the Hamiltonian $H(I,\theta,t) = H_0 + \epsilon \omega_0^2 \cos{\theta}
\cos{\Omega t}$, where the unperturbed  system 
is a pendulum with $H_0 = I^2/2 - \omega_0^2 \cos \theta$ and
$w = H_0/\omega_0^2 - 1$ is the relative variation of
the unperturbed pendulum energy. 
The second line describes the very slow rate of Arnold 
diffusion \cite{arnold}, 
a universally feature of such chaotic separatrix layers
in systems with more than two degrees of freedom \cite{1969,gadiyak,1979}.
More recent results on chaos in separatrix layers and 
Arnold diffusion can be found in \cite{lazutkin,vivaldi,vech1},
see also the book \cite{lich}.

\indent
{\it V.} Here the first equation describes the quantum kicked rotator, 
introduced \cite{kr}, which is the quantized version of the Chirikov 
standard map (see {\it III}).
The classical limit corresponds to $T \propto \hbar \rightarrow 0$,
$k \propto 1/\hbar \rightarrow \infty$ and $K= k T = const$
with $I = T n$. The map gives the evolution of the
wave function after one period of the perturbation 
($\hat{n} = -i \partial/ \partial \theta, \hbar =1$).
The numerical results obtained in \cite{kr} showed that
in the regime of strong chaos $(K \gg 1)$ the rotator energy, 
or the squared number of excited quantum levels $(\Delta n)^2$, 
grows diffusively in time as in the corresponding classical 
system, but only up to a break time $t^*$. 
After this time, the quantum energy excitation is suppressed 
while the classical one continues to diffuse. 
It was shown that $t^*$ grows with $k$, but an  explanation 
of this phenomenon was found only later (see {\it VIII}).

\indent
{\it VI.} The result obtained in \cite{ym} showed
that in general the dynamics of classical Yang-Mills fields is
not completely integrable and can be chaotic. These studies were done
for spatially homogeneous models of Yang-Mills fields introduced 
in \cite{matinyan}, which can be described by an effective 
Hamiltonian with few degrees of freedom $N$. 
For a concrete case with $N=3$ it was found that the dynamics of 
color fields with energy $H$ is characterized by a maximal 
Lyapunov exponent $\Lambda_m \approx 0.38 H^{1/4} > 0$. 
As a result, the Kolmogorov-Sinai entropy $h$ is also positive, 
and the color field oscillations are chaotic. Later, low-energy 
chaos was also found for massive Yang-Mills fields \cite{higgs}.

\indent
{\it VII.} The results \cite{rec} showed  that
the statistics of Poincar\'e recurrences $P(\tau)$ in Hamiltonian 
systems with divided phase spaces decays as a power of time $\tau$ 
with the exponent $p \approx 1.5$. This algebraic decay originates 
from the long sticking of a trajectory near stability islands and 
the consequent slow diffusion on smaller and smaller scales of the phase space. 
This result also implies a slow decay of the correlation functions 
$C(\tau)$ related to recurrences as $C(\tau) \sim \tau P(\tau)$ 
\cite{chi83,karney,phd84}.
The same exponent $p\approx 1.5$ was observed in other maps
and Hamiltonian flows (e.g., the separatrix map, the Chirikov 
standard map etc.), up to times that are about $10^6$ times 
longer than an average return time \cite{karney,phd84,dubna}. 
For larger times the exponent reaches its asymptotic value $p=3$ 
\cite{poster} determined by the scaling properties of the diffusion rate 
near a critical KAM curve \cite{chi83,dubna}. 
Since asymptotically $p>2$, the diffusion rate $D_c$ determined by such 
dynamics remains finite $(D_c \sim \int C(\tau) d \tau < \infty)$ .

\indent
{\it VIII.} In \cite{sov} it is shown that the break time $t^*$,
at which the quantum suppression of classical chaos takes place, 
is proportional to the classical diffusion rate $D$. 
For the kicked rotator, this time scale determines also the number 
of excited unperturbed states 
$\Delta n \sim t^* \sim D \sim k^2  \propto 1/\hbar^2$.
The time scale $t^*$ is much longer than the short Ehrenfest 
time $t_E$ on which a minimal coherent wave packet spreads 
over a large part of the phase space, due to the exponential
local instability determined by the Kolmogorov-Sinai entropy $h$
\cite{zasl,sov}. The analogy between the quantum suppression of 
chaos and the Anderson localization in a disordered 
one-dimensional potential was established in \cite{fishman}.
In this sense the kicked rotator represents the first example
of dynamical localization of chaos in a deterministic system without
any randomness. The localization length is given by $l \sim \Delta n \sim D$.
Recently the dynamical localization in kicked rotator was
observed in the experiments with cold atoms
in a laser field \cite{raizen}. More details on the kicked 
rotator can be find in \cite{sov88,fel90}.

\indent
{\it IX.} On the basis of the results obtained for the kicked 
rotator (see {\it V, VIII}) the dynamical localization length was
found for highly excited hydrogen atoms in a microwave
field, that afforded a determination of the quantum delocalization
border $\epsilon_q$ above which $(\epsilon_0 > \epsilon_q$)
ionization takes place \cite{hatom}.
Here the microwave field strength $\epsilon$ and frequency $\omega$ 
are measured in rescaled atomic units so that
$\epsilon_0 = \epsilon n_0^4, \;\; \omega_0 =\omega n_0^3$, where
$n_0$ is the principal quantum number. 
For $\omega_0 > 1$ the border $\epsilon_q$ can be larger than
the classical chaos border $\epsilon_c = 1/49 \omega_0^{1/3}$ \cite{ufn}.
As a result, for $\epsilon_q > \epsilon_0 > \epsilon_c$ the classical 
atom is completely ionized while the quantum is not. 
Moreover, if in the experiment the frequency $\omega = const$, 
then the quantum border {\it grows} with the level number $n_0$. 
This behavior, predicted in \cite{hatom}, was observed in laboratory
experiments with hydrogen and Rydberg atoms \cite{koch,bayfield,walther}.
As shown in \cite{dec2}, the dynamical localization for
hydrogen atoms in a microwave field can be locally described
by the kicked rotator.
More details about quantum chaos in the microwave ionization
of atoms can be found in \cite{dec1,dec2,dec3}.

\indent
{\it X.} This result \cite{comet} shows that the dynamics of the
Halley comet can be modelled by a simple map (see the Figure):
the comet energy change $(\bar{w} - w)/2$ is a periodic
function of perihelion passage time $x$, measured in 
periods of Jupiter (first equation);  the successive passage 
time $\bar{x}$ is given by the Kepler law (second equation)
[in principle, Saturn also influences the comet's motion \cite{comet}].
This map is approximate but it describes the comet dynamics with 
very high accuracy \cite{comet}.
Studies of this map showed that the Halley comet 
moves chaotically (the KS entropy is positive), and that the 
time of its diffusive escape from the solar system (both forward
and backward in time) is rather short $t_D \approx 4 \times 10^6 years$.
All this information was obtained from only 46 numbers, the 
perihelion passage times, found by extensive numerical simulations 
of other groups and astronomy observations found in historical records.

\hskip 6.0cm {* * * * * * * * * *}

These Chirikov Chaos "Commandments"$^*$, as well as his other results 
(e.g. \cite{faccel,attr,catoms}), are directly related to the modern
physical developments of chaos presented at the Conference.

This Conference attracted more than one hundred participants.
The majority were young researchers from all over the world, 
showing that this branch of science continues to generate
enormous interest.
The Conference logo, representing an eigenstate of the kicked 
rotator (see \cite{maspero} for details), was chosen to emphasize 
the beauty and complexity of chaotic behavior.
The following topics were in the center of discussions of participants:

{\it Nonlinear systems and classical chaos.} This topic was mainly
addressed on the first day of the Conference. The problem of
interaction of nonlinear resonances and their overlap in maps
with quadratic nonlinearity was discussed by Chirikov and Lichtenberg
(related to I, III). The properties of transport 
in quasi-periodic media were discussed by Sinai. 
The recent mathematical results for discrete breathers in
nonlinear lattices were presented by Aubry and MacKay.
The problems of space-time chaos, chaotic patterns
and dynamical phase transitions in extended dynamical
systems were analyzed by Bunimovich, Pikovsky and Ruffo
(partially related to II). The properties of chaos borders and the statistics
of Poincar\'e recurrences in area-preserving 
maps were discussed by Laskar, Artuso and Zaslavsky (see also III, IV, VII).
The special role of symmetric periodic orbits and chaotic dynamics of classical
wave fields were discussed by Seligman and Guarneri.
The chaotic properties of billiards were analyzed by Bunimovich and Mantica.
The problem of hamiltonian round-off errors and
discretization in dynamical systems was highlighted by Vivaldi.

{\it Spectrum, eigenstate properties, quantum ergodicity and localization.}
This topic was represented by investigations of a variety of models.
Multifractal
properties in models of quantum chaos and their relation to wave packet
spreading were addressed by Geisel and Guarneri. The problem of emergence
of quantum ergodicity in billiards was discussed by Borgonovi, Casati, Frahm,
Prange, Prosen, and Ree. Fractal conductance fluctuations in billiards
were studied by Ketzmerick.
The localization of eigenstates
in dynamical models and band random matrices was studied by Fishman, and
Izrailev (related to V, VIII). 
The appearance of deviations from the random matrix theory
and intermediate level statistics was discussed by Bogomolny.
Quantum chaos in open systems,  non-hermitian matrices and 
chaotic scattering 
was addressed by Casati, Maspero, and Fyodorov.
The results of these studies found their applications in experiments with
multimode optical fibers as discussed by Doya, Legrand, and Mortessagne.
The theory of wave-chaotic optical resonators was developed 
and applied to droplet lasing experiments by Stone.

{\it Periodic orbits and quantum chaos.}
New results on periodic-orbit theory for dissipative quantum dynamics were
presented by Haake. Periodic orbits and their signatures in tunnelling,
diffusion and scars were discussed by Creagh, Tanner and Borondo.
Manifestations of periodic orbit quantization in such
mesoscopic systems as a resonance tunneling diode in a tilted
magnetic field were highlighted by Stone.

{\it Quantum chaos in atomic physics.}
An experimental study of dynamical localization 
in the kicked rotator realized with ultra-cold cesium atoms
in a laser field was presented by Raizen, who also
discussed the effects of noise (related to V, VIII).
Experimental investigations of quantum resonances in 
hydrogen atoms in a microwave field and effects 
beyond one dimension were presented by Koch (related to I, IX).
A theory of microwave excitation of chaotic Rydberg atoms in a magnetic
(or static electric) field, where more then 1000 photons are required 
to ionize one atom, was presented by Benenti (see IX).
The chaotic dynamics of quasiparticles in trapped Bose condensates
was studied by Fliesser and Graham.
Properties of stable two electron configurations in strongly
driven helium were discussed by Buchleitner. The statistical
theory of dynamical thermalization and quantum chaos in
complex atoms was presented by Flambaum (related to \cite{catoms}).
A theory of line broadering in recent experiments
with a  gas of interacting cold Rydberg atoms was developed by Akulin.

{\it Quantum chaos in many-body systems.}
This is a relatively recent topic in the field of quantum chaos.
New results for various systems were presented.
The statistics of energy fluctuations of non-interacting fermions
was discussed by Bohigas. Effects of interaction in finite Fermi 
systems were analyzed by Flambaum. Georgeot presented the conditions
for applicability of random matrix theory to quantum spin glass clusters.
The problem of two interacting particles propagating in a random potential
was investigated numerically by Diaz-Sanchez. A transition
from integrable to ergodic dynamics in many-body systems was investigated by
Prosen. A review on quantum chaos in QCD vacuum was given by Shuryak.

The Conference demonstrated how varied are the physical 
applications of the ideas of classical and quantum chaos, 
ranging from QCD \cite{shuryak}, cold \cite{raizen},
Rydberg \cite{koch} and complex \cite{flam} atoms
to mesoscopic physics and chaotic light 
in droplets and microdisk lasers \cite{stone}. 
The achievments of physicists are complementary 
to the impressive mathematical
developments of many twentieth-century mathematicians
which are part of the legacy of Henri Poincar\'e, 
who discovered the first manifestations of what is now called 
deterministic chaos. 
Further mathematical results 
in this area can be found in the
parallel special issue of the Annales de l'Institut  Henri Poincar\'e 
on ``Classical and Quantum Chaos'', also dedicated to Boris Chirikov
on the occasion of his 70th birthday \cite{ihp}.

In closing, we would like to thank the Conference sponsors,
who helped to make this event possible:
CNRS, Universit\'e Paul Sabatier, Institut Universitaire
de France, IRSAMC, R\'egion Midi-Pyr\'en\'ees.
We owe special thanks to Physica D, the journal where Boris 
Chirikov and his friend Joe Ford collaborated during many years. 
Our warmest thanks go to Sylvia Scaldaferro and Robert Fleckinger,
for their day-by-day assistance in the organization; we also 
thank Armelle Barelli, Klaus Frahm and Bertrand Georgeot.
We are thankful to F.M.Izrailev, V.V.Vecheslavov
and F.Vivaldi for their friendly remarks we used here.
Finally, we express our gratitude to the invited speakers 
and to the authors of the contributions to this volume,
who enthusiastically accepted the invitation to participate.

\begin{center}
{\large J.~Bellissard, O.~Bohigas, G.~Casati 
and D.~L.~Shepelyansky} \end{center}
\begin{center}
{Toulouse, 21 September, 1998 } \end{center}

\newpage

\begin{center}
{\large \bf  Remarks on Boris (a Russian) by Peter (an American)\\
 in Toulouse (France): \\
Yes, physics is international ! }

{ Peter M. Koch}

{ Department of Physics and Astronomy\\
State University of New York\\
Stony Brook, NY 11794-3800, USA}

{(Received 3 October, 1998)}

\parskip 0.3cm

{\bf Abstract}

{ The author was deeply honored to be requested to give the after-dinner speech 
at the banquet held on 17 July 1998 as part of the conference "Classical 
Chaos and its Quantum Manifestations" in honor of the 70th birthday of Boris 
Chirikov.  An unusual aspect of this after-dinner speech was its being given 
before dinner.  The text has been edited to conform, as closely as possible, 
to the words that were actually spoken.\\
---------}

\end{center}

A reporter once asked the elderly Winston Churchill of Great Britain what in 
life was the most difficult test.  Churchill replied, "To climb a ladder 
leaning towards you, to kiss a girl leaning away from you and third, to give 
an after dinner speech."  

I avoid ladders leaning toward me.  So far this evening, the opportunity to 
kiss a girl, leaning whatever way, has not presented itself.  So my test is 
the third.  It was an honor to be asked to address you after tonight's lovely 
banquet, though the organizers have just requested that I do so before dinner.  
Throughout the meeting, and especially tonight we join in celebrating the 70th 
birthday and science of our dear colleague and friend Boris Chirikov.  

Most of you, as Boris's former students, collaborators, or friends could have 
been asked to speak tonight and, I'm sure, would have gladly agreed.  Perhaps 
it's a bit strange  that I, an American, have been recruited to celebrate 
here, in France, Boris, a Russian, before you all, who come from various 
corners of the world.  

Ya na gaviru parruski, so I cannot honor Boris in his native language... 

which reminds me of the story of two sheep who were grazing in a pasture.  The 
first sheep said, "BAAAA."  His companion looked up and replied, "MOOOO."  The 
first sheep was rather surprised by this and asked, "What are you doing?  
Sheep don't say MOOOO; they say BAAAAA."  "I know," replied his friend.  "I 
was just practicing a foreign language."  

So if I'm saying "MOOOO" where many of you would expect to hear "BAAAA", 
please bear with me.  

For many people, birthdays are like glasses of wine: after you've had a few,  
you don't bother to count them.

But Boris gives us a way to measure his years: by his scientific development 
as a young researcher and later professor and mentor to his students and 
colleagues, by the productivity marked by his many important papers, and by 
the rich stories of his life.  If I don't get them all without small errors, 
please forgive me.  

Just a few facts.  His birth was in Orel on the 6th of June 1928, so his 
boyhood came in the internal turbulence of the Soviet Union of the 1930's.  
With the father absent from their small family, he and his mother fled famine 
and went around 1936-7 to relatives in Leningrad, where they, of course, found 
more difficult times after the outbreak of war in Europe.  In 1941 or 42 they 
were evacuated from Leningrad to the northern Caucasus, a region that about 4 
months later was conquered and occupied by the German Army.  

How many of you in the audience now have teen-aged children?  (A show of hands 
indicated that many did.)  My son Nathan is 13, the same age as Boris in 
1941-2.  Despite the trials of adolescent life in America, I know that my 
Nathan at 13 faces nothing like the wartime dislocations that came to Boris 
and so many others during the terrible wartime.  Children should never have to 
face such things, but we still see around the world that they do.  Let us all 
pray that human beings can learn to put such behavior behind them.  

Less than a year later, the counter offensive of the Soviet Army moved the 
frontier westward again and liberated the region, but Boris soon faced sadness 
again.  After a protracted illness, Boris's mother died around 1944.  He was 
now an orphan.  Fortunately for his future and ours, a teacher at his school 
took him into her home and life went on.  

In fact, it was during this period near and just after the end of the war that 
Boris became the sweet man we have all come to know.  

Sweet, you say?  Well many years later, Peter Scherer, a young German 
physicist, recognized Boris's sweetness and coded it into the caricature he 
drew of Boris at the Les Houches theoretical summer school in 1989.  Though I 
was not there, I know that a number of you here were.  (The only transparency 
used in the talk was then shown.  It was the caricature of Boris appearing in 
Chaos and Quantum Physics, M.-J. Giannoni, A. Voros, and J. Zinn-Justin, 
editors (Elsevier, Amsterdam, 1991, page 444).)   

We all recognize Boris lecturing.  But notice Boris's small listener, buzzing 
around feverishly.  Maybe the question marks mean that this bee is not 
sufficiently educated, but the bee is certainly not stupid.  This bee knows 
Boris's sweetness, for bees are drawn to sugar.  

What does sugar have to do with the picture, you ask?  Well, what you may not 
know is that from about 1944 to 1947 Boris's place in socialist labor was 
working in a sugar factory.  Boris became quite the expert on sugar, and knows 
good and bad sugar when he tastes it.  

(The speaker turned to Boris.)  With that the case, Boris, let me present you 
with a small present.  Please accept this small box of American sugar, but I 
do ask that after you try it, you do make public your referee report.   

Let me jump forward tell you the story of how in 1987 I met Boris in his 
native land.  In December 1986 I had been invited by telex (these were the 
days before widespread email and the Internet) to be a speaker at the IXth 
Vavilov Conference on Nonlinear Optics in Novosibirsk in mid June 1987.  For 
this I was to be, for about two weeks, an official guest of the Soviet Academy 
of Sciences.  Around that time it was also arranged that there would be a 
small workshop about two weeks earlier in Riga, then in the Latvian SSR.  
Therefore, my invitations to these meetings meant that I would change "status" 
while in the USSR, from an "ordinary" foreign person for the first week, 
around the Riga workshop, to an important guest of the Academy, for the last 
two weeks or so around the Novosibirsk meeting.  

Well, Latvia was nice, the workshop hosted by Robert Damburg was small and 
wonderful, and there I met Boris as well as Dima, our co-host this time 
around.  

After the Riga workshop, I underwent in Moscow the transformation from 
ordinary American to official guest of the Soviet Academy.  A chauffered car 
was made available to take me from lab to lab.  Young scientists were 
assigned to shepherd me around.  Emboldened, I decided to make a firm request.  
I wanted to travel to Novosibirsk by train, not by the Aeroflot flight already 
arranged.  

No, no, they said.  It's too far.  Do you know how far it is, they asked?  
Sure, I said.  About the same as from New York City to Denver, which in the 
USA I would never do.  I'm too busy!  It takes too long!  But I have the time 
here, and I want to see your big country.  

But we've no one to go with you, they said.  No problem, I said.  I'll be 
fine.  

But you don't speak Russian, they said.  True, I said, but I'll be OK.  

Finally, my polite firmness won out, and the train travel was booked.  On 
train number 7, the same tracks as the Trans Siberian Railway, I'm told.  A 
soft car, with me alone in the compartment.  A kind young Moscow physicist 
donated a copy of a Russian phrase book prepared in English for the Moscow 
Olympic Games, which the USA and some other countries had boycotted.  I 
suppose that meant that lots of phrase books were left over.  

Anyway, off I went.  The view out the window of my train compartment was 
wonderful, and at mealtimes I did manage to find the dining car, where the 
kind people really looked after me.  53 hours later (I think it was), on a 
Sunday evening, I arrived in Novosibirsk.  This turned out to be quite an 
event.  First, no one told me that the Soviet trains traveled on Moscow time!  
According to the time schedule on the wall of the train, I thought I would be 
arriving in the afternoon in Novosibirsk.  Out on the train station platform, 
I noticed that the sun was suspiciously low in the sky.  It was definitely not 
afternoon.  And the big digital clock on the Novosibirsk hotel did not say the 
time that I expected.  It said 20 hours as I recall.  

Moreover, as all the passengers leaving the train walked away on the platform, 
it became clear that no one was at the station to meet me.  

What to do?  

Well, having been in the USSR already for nearly two weeks, I now recognized 
the uniforms.  I walked up to a blue one, a Militsia officer, cleared my 
throat, and said, "Amerikanskii.  Ya na gaviru paruskii."  What a look on his 
face.  I'll never forget it!  He started to speak to me in Russian, but then, 
realizing the futility, he stopped.  

I showed him the one piece of official paper I had with Cyrillic letters on 
it.  It was a letter written in English from the conference organizers, but it 
was on the official letter paper of the Institute of Thermal Physics (Institut 
Teplofiziki).  The Militsia officer took my precious piece of paper, read the 
Russian top of it, and evidently got his plan.  He took me to the head of the 
taxi queue, spoke to the taxi master, and got me put in a taxi to go to 
Akademgorodok.  The taxi driver had my precious piece of Russian paper.  

Just as the sun was nearly setting, we arrived in Akademgorodok.  Of course, 
the taxi man took me to the institute.  That was what was printed on the top 
of my letter.  But it was Sunday night.  The institute was closed!  Which the 
taxi driver soon discovered after getting out of the taxi and knocking on the 
front door.  He returned to the taxi, looked at me, and shrugged his 
shoulders, as if to say, "What now, Amerikanskii?"  

Something made me think of looking in my American "Fodor's Guide" to the 
Soviet Union.  Listed under Akademgorodok was the name of only one hotel.  
I said this to the taxi driver.  He asked some people walking by where it was, 
and then quickly drove to it.  It was only a block or two away.  

On the outside of the hotel, I could see in English a sign that said something 
like "Welcome to the IXth Vavilov Conference".  We had found the right place.  

After a good night's sleep in my room, I was in the hotel lobby the next 
morning when all the Soviet scientists from Moscow showed up after their very 
early morning flight from Moscow.  I don't recall who it was, but when one of 
them saw me sitting there in the lobby, all the color drained out of his face.  
He came up to me and said, "You're here!  You weren't supposed to arrive until 
today!  How did you get here?  No one was here to meet you!"  

I just smiled and said, "It's a long story, but I got here just fine last 
evening."  

A couple of other Moscow scientists recognized me and came over to talk.  I 
remember clearly one asked me, "You took the train all the way from Moscow.  
How was it?"  I said, "It was wonderful."  He said, "Really?  I've always 
wanted to do this!"

That is how I got to Novosibirsk in the early days of glasnost and 
perestroika.  

Now two short stories from the Vavilov Conference, one about Dima and one 
about Boris.  

The Plenary Talks were held in the Large Auditorium, where earphones were 
supplied at every seat for simultaneous translation (just like the United 
Nations).  If you spoke in Russian, it was translated to English, and vice 
versa.  Well done, too.  The translators must have known some science because 
they seemed very good to me.  

The Seminar talks that were held in the evening, upstairs in smaller rooms, 
were different.  I still have the conference program, and the following 
occured on Thursday evening in the seminar "Rydberg states and strong field".  
The session chairman, listed in the program as I.M. Beterov, opened the 
session in English and asked all speakers to give their talks in English 
because of the Americans and other foreign scientists present, even though we 
were fractionally small in number.  Well over 90\% of those in the room were 
Russian speakers!  In the middle of Siberia, I was surprised.  

The first speaker was Dima Shepelyansky, and as always, he did very well, both 
with the science and with the English.  The second speaker was Nikolai Delone, 
from Moscow, who speaks French well but not English.  He said a few words in 
Russian to Dima, and Dima announced, in English, that the speaker, Delone, 
would speak in Russian but had asked Dima to translate into English.  Dima had 
evidently not expected this, but he agreed to do so.  

I remember Nikolai's talk well.  He used lots of words but only one 
transparency.  After every minute or so of talking, he would stop for Dima to 
translate.  Dima would say, "The speaks says blah blah blah ... ."  Dima was 
evidently struggling, not with the English but to understand just what the 
science was he was supposed to be translating.  During the third or fourth 
chunk of Russian that Nikolai presented, I saw a pained look come on Dima's 
face.  It was clear why Dima looked this way, because he opened the 
translation of this part by saying, "The speaker says, but I do not agree with 
this part! ... ."  The words coming out of Dima's mouth were Nikolai's, but 
Dima could not agree with them and wanted to make sure that the audience knew 
this.  

As I recall, it was the next Russian speaker who started his talk with a few 
words of English that said, "I am Russian.  I do not speak English well.  I 
will not give my talk in English.  I will give my talk in Russian!"  He must 
have said this in Russian immediately after this, because these words brought 
a cheer from the audience.  Another Russian Revolution!  But this time, 
no fighting nor bloodshed.  

As it happened, I was sitting next to Boris at this seminar, so Boris just 
leaned over to me and whispered in my ear, "I will translate for you."  And so 
he did, most skillfully summarizing in a few words the important point after 
each few sentences of the speaker.  

I don't recall if it was our revolutionary speaker or the next Russian one 
when the following happened.  The speaker began, in Russian, and Boris started 
translating for me.  After a few minutes into the talk, Boris leaned over 
especially close and whispered to me something like, "The speaker is very 
confused and doesn't understand what he is talking about, so I will stop 
translating what he is saying."  

Now, let me tell you, this is an efficient way to attend a seminar: not only 
simultaneous translation, you also get an analog filter on the science!  Boris 
is not one to avoid voicing his opinions. 

Nor is Boris one to avoid making insightful decisions, as the early part of 
his career shows.  We know Boris as a theorist, indeed the Director of the 
Theoretical Division of the Institute of Nuclear Physics.  But I find it 
fascinating that he started his career as an experimenter!   

In his 5th year of studies in what became the Moscow Physicotechnical 
Institute, he was a student-apprentice researcher in the Heat Engineering Lab, 
now called the Institute of Theoretical and Experimental Physics.  After 
graduation around 1952, he was invited to remain there.  His project involved 
a Wilson cloud chamber for high energy particle tracking and identification.  
You recall, photographs were taken of tracks of droplets that formed in the 
supersaturated vapor after the passage of ionizing radiation.  The photographs 
had to "scanned", work that was tedious and time-consuming for the girl 
technicians.  Boris's simple, time-saving proposal was they should count the 
number of gaps.  This was statistically related to the number of drops, but it 
could be done more quickly.  
 
As Boris told us yesterday, he agreed to transfer in 1954 to what was later 
called the Kurchatov Institute of Atomic Energy.  Actually, he was recruited 
there by the theorist Andrei Mikhailovich Budker, who had earlier taught the 
student Boris.  

With his first research student Volosov, Chirikov did crucial experiments on 
the limiting current of electron beams.  It was here that Boris's interests
in nonlinear phenomena and stochastic processes began. 

In 1958 Budker was selected to form the new Institute of Nuclear Physics at 
the new Akademgorodok (Academy City) being constructed outside Novosibirsk, 
and the actual move there two years later took the team, including Chirikov, 
to Siberia.  

Boris first presented his results on the stochastic instability of 
magnetically confined plasma at the Kurchatov seminar in Moscow in 1958, when 
the plasma research was classified secret.  Only after the London plasma 
conference of 1958 did the results become public, and Kurchatov ordered the 
plasma results to be published quickly.  This led to Boris's celebrated 1959 
theoretical paper in a special issue of the journal Atomic Energy.  Boris had 
started his career as an experimenter, but the world would now know him as the 
theorist who invented the resonance overlap criterion.  

What you may not know is the story of the writing of the paper in the same 
journal issue that describes the related plasma experiments of S. Rodionov.  
Though Rodionov's name appears as the sole author, the paper was written by 
Chirikov.  

Why?  The story goes that Rodionov had broken his right hand (probably during 
skiing) and was in the hospital.  Boris was ordered by the KGB to take his 
secret notes, go to the hospital, and write the paper from the words of 
Rodionov.  The KGB orders included that Boris take a weapon, a revolver, to 
ensure the security of the secret documents, but Boris refused, arguing that 
it would be too dangerous to take a revolver on the public buses that, in 
those days, were always very overcrowded with people.  Finally, the KGB agreed 
that Boris would not have to carry the revolver, but he was oblighed to return 
all his notes, including the "Rodionov manuscript" back to the secure place.  

What we now all know as the "Chirikov resonance overlap criterion" came as a 
result of Boris's generalizing the theoretical analysis he had first performed 
for the stochastic instability of confined plasma.  Apparently, the first full 
experimental confirmation of Boris's criterion came at the end of the 1960's, 
with experiments in Novosibirsk on circulating electron beams.  

Chirikov's later widespread and continuing interactions with Western 
scientists was certainly stimulated by the pioneering results of Chirikov 
that, fortunately, were published in the open literature and that I have 
decribed briefly.  However, how each of Boris's personal relationships with 
Western scientitsts began and developed depended, of course in those times of 
Cold War, on the occasional interruption by more openness.  

Boris's long association with Joe Ford of Atlanta, USA, began when Boris and 
Joe met at a conference in Kiev in 1966.  We are all sad that Joe is no 
longer with us and cannot be here.  

Alerted by Ford, Giulio Casati from Milano visited Novosibirsk in 1976 and 
began a long collaboration with Chirikov and his students that continues up to 
the present day.  As we all know, this circle even widened to the younger 
associates of Ford in the USA and Casati in Italy.  

Let me begin to close by reading some of the reminiscences sent to me on July 
3rd by another of Boris's many friends and collaborators, Andy Sessler of the 
Lawrence Berkeley Laboratory in the USA.  As it turns out, Andy is, this year, 
President of the American Physical Society, but it is clear that his remarks 
are of a personal nature.  

I shall read from Andy Sessler's remarks, a copy of which I have just given 
to Boris:

     I first met Boris in March of 1965 when Budker invited a small number of 
     people (about 12) to Novosibirsk to discuss the technical aspects of 
     storage rings. At that time he told me about his new work, which was 
     unpublished (I believe) at that time and has subsequently become known as 
     the `Chirikov criteria'. 

     We `hit it off together' and ever since then have been good friends. At 
     that time we did a number of `fun things' like spending evening Under The 
     Integral Sign (a scientific club; really an eating club and night club) 
     and also going cross country skiing. To do that one checked out skis, of 
     course free in those Soviet Days, and that was done one afternoon. Then, 
     the next day, we went to the ski area. The ski area was a small hill 
     (this was Siberia) covered with pine trees. The Soviets could change 
     direction while going down hill, but the Americans were doing their best 
     just to stand up on their skis. That meant that in order not to hit a 
     tree you had to point your skis correctly, before you started down the 
     hill, to about a milliradian. I was doing that fine through most of the 
     afternoon, but then I mis-calculated and hit a tree. I broke the tip of 
     the ski off (and was damn fortunate not to have broken anything else) and 
     remember walking, through deep snow, for what seemed like miles and 
     miles. For many years I had the ski tip as a souvenir of my first meeting 
     of Boris.

     Some time later, in 1967, I was spending the year at CERN and Boris 
     visited us. My chance to get even. I suggested that we go to Zermat and 
     do a bit of real (down hill) skiing. So my family (5 of us) and two 
     Soviets, Boris and Ben Sidorov (now deputy director of the Budker 
     Institute), piled into my car and drove from Geneva to where one takes 
     the train to Zermat. 

     The next day was terribly cold and everyone, except Boris, decided not 
     to try and ski. Boris was not going to miss out on anything and I, as 
     host, felt I must go with him. Me in lots of down and him in a simple 
     sweater. Well, it was really cold. We rode the lift up, skied down and 
     when I took off my gloves my fingers were all white. Boris rushed me to a 
     first aid station and proceeded to rub snow on my hands.  Well, he saved 
     my fingers, sent me in for the day, and continued to ski all day, coming 
     in, at the end of the day in fine form. 

     Through the years we continued to send cards  (as well as scientific 
     papers) and I remember one where Boris said it was 40 below and he had 
     stopped skiing. Not to be out done, I sent back a card saying that the 
     Soviets might stop at 40, but Americans certainly kept skiing. He then 
     wrote back saying that I didn't understand: it was the skis that stopped 
     working when it got so cold."
 
     Boris and I, did, once write a paper together. Well, Boris really did all 
     the work, but I do remember a very pleasant day working--for some reason--
     in his kitchen. Boris had the idea that there hadn't been a paper since 
     World War II co-authored by a Russian, German, and American. (I don't 
     know if this was true or not, but it was an interesting thought.) So, we 
     invited Eberhard Keil into the collaboration."

     Once, Boris's wife, Olga, was `allowed', I think that is the right term, 
     to go on a vacation consisting of a cruise on the Black Sea (and maybe 
     also the Mediterranean). The cruise was for artists (she was a well-known 
     opera singer). The first leg consisted of air to Moscow and it was 
     arranged that she and I went together. In Moscow she escorted me around, 
     including a very lengthy tour--and very special tour led by a friend of 
     hers--to the Tretyahov Gallery. All very good, but she didn't know a word 
     of English and I don't know a word of Russian; we just smiled at each 
     other for a few days."

     Boris, I hope you have a great time at this Conference at which your 70th 
     birthday (how can we all be so old so soon?) is properly noted. The honor 
     is richly deserved. I feel touched to have had my life touched by you."

     ----- Andy Sessler

And now, I will close my talk directly to Boris:

     Boris Valerianovich, my vashi druzya zhelaem vam prodolzhat' prodvigatsya 
     po puti resheniya fundamentalnykh problem.

                          Acknowledgements 

I greatly appreciate the help given by Dima Shepelyansky and Edward Shuryak 
for preparation of this talk.  Andy Sessler kindly furnished his recollections 
and gave permission to use them in the oral presentation and in the written 
text.  

\newpage
\begin{center}
{\large \bf List of Contributions to the Volume }
\end{center}

1. {\it Sputnik of Chaos (Intoduction)}\\ by J.~Bellissard,
   O.~Bohigas, G.~Casati and D.~L.~Shepelyansky
   
2. {\it Threshold to Global Diffusion in a Nonmonotonic 
   Map with Quadratic Nonlinearity}\\ by G.~Corso and
   A.~J.~Lichtenberg
   
3. {\it  Forcing Oscillatory Media: Phase Kinks vs. Synchronization} \\
   by  H.~ Chat\'e, A.~S.~Pikovsky and O.~Rudzick
  
4. {\it Space-Time Chaos in Spatially Continuous Systems}\\
   by L.~Bunimovich
   
5. {\it Chaos and Statistical Mechanics in the Hamiltonian 
   Mean Field model}\\
   by V.~Latora and  S.~Ruffo
   
6. {\it Chaotic layer of a  nonlinear resonance
   driven by quasiperiodic perturbation}\\
   by V.~V.~Vecheslavov
   
7. {\it Correlation decay and return time statistics}\\
   R.~Artuso
   
8. {\it Polygonal Billiards Revisited: A Model 
   of Quantum A-Integrability}\\
   by G.~Mantica
   
9. {\it Quantum chaos with cesium atoms: pushing the boundaries}\\
   by B.~G.~Klappauf, W.~H.~Oskay, D.~A.~Steck and M.~G.~Raizen
   
10. {\it Beyond (1d + time) dynamics in the microwave ionization of
excited H atoms: Surprises from experiments with collinear static and linearly
polarized electric fields}\\
   by P.~M.~Koch, E.~J.~Galvez and S.~A.~Zelazny

11. {\it Stable classical configurations in strongly driven helium}\\
    by P.~Schlagheck and A.~Buchleitner
   
12. {\it Level-band problem and many-body effects in cold Rydberg atoms}\\
    by V.~M.~Akulin, F. de Tomasi, I.~Mourachko and P.~Pillet
    
13. {\it Classical quasiparticle dynamics and chaos in trapped Bose
    condensates}\\
    by M.~Fliesser and R.~Graham
    
14. {\it Finite-length Lyapunov exponents and conductance for quasi-1D
    disordered solids}\\
    by T.~Kottos, F.~M.~Izrailev and A.~Politi
    
15. {\it  Models for Chaotic and Regular Motion on the Fermi Surface}\\
    by A.~Iomin and S.~Fishman

\newpage   
16. {\it On the distribution of the total energy of a system of
    non-interacting fermions: random matrix and semiclassical estimates}\\
    by O.~Bohigas, P.~Leboeuf and M.~J.~Sanchez
    
17. {\it Quantum chaos in many-body systems: What can we
    learn from the Ce atom?}\\
    by V.~V.~Flambaum, A.~A.~Gribakina, G.~F.~Gribakin and
    I.~V.~Ponomarev
    
18. {\it Quantum Chaos in Quantum Wells }\\
    by E.~E.~Narimanov  and  A.~D.~Stone
    
19. {\it Efficient Diagonalization of Kicked Quantum Systems}\\
    by R.~Ketzmerick, K.~Kruse and T.~Geisel
    
20. {\it On the Special Role of Symmetric Periodic Orbits in a 
     Chaotic System}\\
     by L.~Benet, C.~Jung, T.~Papenbrock and T.~H.~Seligman
     
21. {\it Semiclassics for a Dissipative Quantum Map}\\
    by D.~Braun, P.~A.~Braun and F.~Haake
    
22. {\it The quantum mechanics of chaotic billiards}\\
    by G.~Casati and T.~Prosen  
    
23. {\it Quantum fractal eigenstates}\\
    by G.~Casati, G.~Maspero and D.~L.~Shepelyansky
    
24. {\it Cantori and dynamical localization in the Bunimovich Stadium}\\
    by F.~Borgonovi, P.~Conti, D.Rebuzzi, B.~Hu and B.~Li
    
25. {\it Remarks on Boris (a Russian) by Peter (an American)
    in Toulouse (France): Yes, physics is international !
    (unformal Conclusion)}\\
    by P.~M.~Koch

\end{document}